# The structural architecture of the Los Humeros volcanic complex and geothermal field

Gianluca Norini[1], Gerardo Carrasco–Núñez[2], Fernando Corbo-Camargo[2], Javier Lermo[3], Javier Hernández Rojas[2], César Castro[2], Marco Bonini[4], Domenico Montanari[4], Giacomo Corti[4], Giovanna Moratti[4], Luigi Piccardi[4], Guillermo Chavez[5], Maria Clara Zuluaga[6], Miguel Ramirez[7], Fidel Cedillo[8]

[1]Istituto di Geologia Ambientale e Geoingegneria, Consiglio Nazionale delle Ricerche, Milano, Italy
[2]Centro de Geociencias, Universidad Nacional Autónoma de México, Querétaro, México
[3]Instituto de Ingeniería, Universidad Nacional Autónoma de México, CDMX, México
[4]Istituto di Geoscienze e Georisorse, Consiglio Nazionale delle Ricerche, Firenze, Italy
[5]Polaris Energy Nicaragua S.A., Managua, Nicaragua
[6]Dirección de Recursos Minerales, Servicio Geológico Colombiano, Bogotá, Colombia
[7]Comisión Federal de Electricidad, Gerencia de Proyectos Geotermoeléctricos SEDE, Morelia, Michoacán, México
[8]Seismocontrol S.A. de C.V., Ciudad de Mexico, Mexico

Corresponding author: G. Norini, tel. +390266173334, fax +390228311442, gianluca.norini@cnr.it

*Abstract*

*The Los Humeros Volcanic Complex (LHVC) is a large silicic caldera complex in the Trans-Mexican Volcanic Belt (TMVB), hosting a geothermal field currently in exploitation by the Comisión Federal de Electricidad (CFE) of Mexico, with an installed capacity of ca. 95 MW of electric power. Understanding the structural architecture of LHVC is important to get insights into the interplay between the volcano-tectonic setting and the characteristics of the geothermal resources in the area. The analysis of volcanotectonic interplay in LHVC benefits from the availability of subsurface data obtained during the exploration of the geothermal reservoir that allows the achievement of a 3D structural view of the volcano system. The LHVC thus represents an important natural laboratory for the development of general models of volcano-tectonic interaction in calderas. In this study, we discuss a structural model of LHVC based on morphostructural and field analysis, integrated with well logs, focal mechanism solutions and magnetotelluric imaging. The structural analysis suggests that inherited regional tectonic structures recognized in the basement played an important role in the evolution of the magma feeding system, caldera collapses and post-caldera deformations. These inherited weak planes have been reactivated by resurgence faults and post-caldera magma-driven hydrofractures under a local radial stress field generated by the shallow LHVC magmatic/hydrothermal system. The local stress field induced caldera resurgence and volcanotectonic faulting. The results of this study are important to better constrain the structural architecture of large caldera complexes. Also, our study is useful to understand the structure of the Los Humeros geothermal field and support the exploration of deeper Super-Hot Geothermal Systems (SHGSs) and engineering of Enhanced Geothermal Systems (EGSs) for electric power production in the LHVC and other active resurgent calderas.*

**Keywords:** volcano-tectonic interplay; caldera collapse; caldera resurgence; geothermal exploration.

## 1. Introduction

The structural architecture of large silicic calderas is strictly related to the interplay among the geological setting of the substratum, the stress field, the dynamics of the magma reservoirs and the eruptive history of the volcanic system (e.g. Lipman, 2000; Cole et al., 2005; Acocella, 2007; Gudmundsson, 2008). Inherited structures, regional stress regime and active tectonic faults in the basement influence the location and geometry of magma chambers and overlying volcanic edifices and calderas (e.g. Robertson et al., 2015; Wadge et eal., 2016). Prior to caldera collapse, the load of pre–existing volcanoes modifies the local stress field and may focus the deformation in the basement, generating volcanotectonic faults with variable geometry and kinematics (e.g. Norini et al., 2010). During the collapse of silicic calderas, the rapid emission of large amounts of pyroclastic material triggers the formation of ring faults, extending from the topographic surface down to a depth of several kilometers in the crust and displacing the roof of the magma chamber (e.g. Lipman, 2000; Cole et al., 2005). The geometry of these faults is influenced by the shape and depth of the empting magma reservoir and may also be controlled by pre–existing steep discontinuities in the crust (e.g. Folch and Marti, 2004; Gudmundsson, 2008; Kennedy et al., 2018). After collapse, changing fluid overpressure in the magmatic reservoirs and associated hydrothermal system may drive faulting and folding of the overlying rocks, with volcanotectonic ground deformations and the resurgence of the caldera floor (e.g. Acocella et al., 2000, 2004; Saunders, 2004).

All the above mentioned tectonic and volcanotectonic structures are common features of active and extinct silicic calderas, having a great influence on their geological evolution, eruptive history, geothermal potential and occurrence of ore deposits. However, few of the deformation features associated with recent and active silicic calderas have a clear surface expression, due to the emplacement of large volumes of young volcanic, alluvial and lacustrine sediments, commonly covering up tectonic and volcanotectonic structures (e.g. Marti et al., 2008; Hutchison et al., 2015). Also, the scarcity of eroded caldera systems in the stratigraphic record, where the geometry and kinematics of the basement structures, magmatic intrusions and volcanotectonic deformation features can be observed, imposes significant restrictions to the study of the geology and evolution of large silicic calderas (e.g. Branney, 1991; John, 1995; Marti et al., 2008; Mueller et al., 2008, 2009). Thus, although much has been done so far, the knowledge of many aspects of the structural architecture of silicic calderas is still poorly constrained by data from natural caldera complexes, and mainly relies on indirect analogue and numerical modeling results (e.g. Acocella, 2007; Hardy, 2008; Geyer and Martí, 2014).

The integration of geologic data and interpretations from field observations, boreholes samples, well logs, geophysical imaging and analysis of active seismicity would be fundamental to better depict the three–dimensional structural architecture of these natural systems. Unfortunately, this amount of expensive surface and subsurface data is rarely available to the scientific community for the study of silicic calderas, as it is usually owned by private energy and mineral enterprises for the exploration and exploitation of geothermal systems and ore deposits.

Recently, an agreement with the Comision Federal de Electricidad (CFE) of Mexico, owning the geothermal exploitation permit, made available subsurface data archives and prompted the acquisition and improvement of geophysical imaging, seismic data, high–resolution digital elevation model (DEM) and geological field data of the Los Humeros Volcanic Complex (LHVC) in Mexico. The LHVC is one of the largest active silicic caldera complex of North America, hosting an active hydrothermal system (e.g. Norini et al., 2015; Carrasco–Núñez et al., 2017a, 2017b, 2018). Since the late '70s, the exploration of the geothermal reservoir of the LHVC, aimed at the production of electric power, generated large amounts of information on the subsurface geology of the central sector of the caldera complex, with the drilling of more than 60 wells up to 3000 m depth (e.g. Cedillo, 2000; Arellano et al., 2003; Gutiérrez–Negrín and Izquierdo–Montalvo, 2010; Rocha López et al., 2010). In the last years, local seismic data and resistivity imaging have been acquired to better constrain the subsurface structure of this geothermal reservoir and the hosting caldera complex (Lermo et al., 2016; Urban and Lermo 2017; Arzate et al., 2018). Also, stereo satellite images at 0.5 m resolution of the LHVC have been obtained to calculate an accurate 1 m resolution DEM, as well as to depict with high precision the geometry of recent deformation features at the surface. Starting from the achievements presented by Norini et al. (2015), based on limited field survey and remote sensing, all the new data and some of the archive subsurface information, coupled with a new extensive structural fieldwork in the LHVC and surrounding basement, have been analyzed and integrated in a 3D Geographic Information System (GIS) to provide a structural model of the volcano. This model considers that the structural architecture of the caldera complex and geothermal field was obtained as a result of the interplay among inherited tectonic structures in the basement and recent volcanotectonic deformation induced by caldera collapse and resurgence. The final aim of our paper is to contribute to a better comprehension of the structure of large silicic calderas, useful for the assessment and management of the volcanic hazard and the exploration and exploitation of the natural resources associated with these volcanoes.

## 2. Regional setting and geological overview of the LHVC

The Trans–Mexican Volcanic Belt (TMVB) is a Miocene–Holocene continental volcanic arc (Ferrari et al., 2012, and references therein). The Pleistocene–Holocene LHVC is the largest active caldera of the TMVB and is located in the eastern sector of the arc (Fig. 1) (Carrasco–Núñez et al., 2017a, 2017b, and references therein). This sector of the TMVB is mainly built on Mesozoic sedimentary rocks involved in the Late Cretaceous–Eocene compressive orogenic phase that generated the Mexican fold and thrust belt (MFTB) (Fitz–Diaz et al., 2017, and references therein). In this area, the MFTB is thought to mainly have thin–skinned structural style above a basal detachment located at about 4–5 km depth over the Precambrian–Paleozoic crystalline basement, made of metamorphic and intrusive rocks, including green schists, granodiorites and granites (e.g. Suter, 1987; Suter et al., 1997; Ortuño–Arzate et al., 2003; Angeles–Moreno, 2012; Fitz–Diaz et al., 2017). Pervasive folding with different wavelengths and thrust faults of the MFTB affected Jurassic-Cretaceous limestone and terrigenous sedimentary rocks, with slices of the metamorphic basement cropping out in the core of the major anticline folds (e.g. Yáñez and García, 1980; Fitz–Diaz et al., 2017). The folded succession has an overall NW–SE strike and NE–directed tectonic transport, although with local trend variations, and has been intruded by Eocene–Miocene granite and granodiorite stocks (e.g. Ferriz and Mahood, 1984; Suter, 1987; Gutiérrez–Negrín and Izquierdo–Montalvo, 2010; Carrasco–Núñez et al., 2017b; Fitz–Diaz et al., 2017).

After the MFTB orogenic phase, the LHVC area underwent a limited Eocene–Pliocene extensional tectonic deformation, associated with scattered, mainly NE–striking normal faults. These extensional structures have been described as preferential pathways for the Eocene–Oligocene magmatic intrusions preceding the onset of the TMVB volcanism and, later, for the emplacement of Plio–Quaternary volcanoes (e.g. Campos–Enríquez and Garduño–Monroy, 1987; López–Hernández, 1995).

The TMVB volcanic activity in the LHVC area started in the Late Miocene (at around 10.5 Ma ago), with the emplacement of the Cuyuaco and Alseseca units, mainly composed of fractured andesites and basaltic lava flows with a cumulative thickness up to 800–900 m (Cedillo–Rodríguez, 1984; López–Hernández, 1995; Yáñez and García, 1982), which correlate with the Cerro Grande volcanic complex, dated between 8.9-11 Ma (Carrasco–Núñez et al., 1997). Volcanic activity also occurred between 5 and 1.55 Ma, with the emplacement of the fractured andesites of the Teziutlán volcanic unit, seldom intercalated with volcaniclastic levels in the upper part. This volcanic unit has a thickness of more than 1500 m, as measured in some of the geothermal wells within the LHVC (e.g. Yañez and García, 1982; Ferriz and Mahood, 1984; Cedillo–Rodríguez, 1997). Recent $^{40}Ar/^{39}Ar$ dating of these units reveals ages varying between 1.46±0.31 Ma and 2.65±0.42 Ma (Carrasco–Núñez et al., 2017a).

The volcanic evolution of the basaltic andesite–rhyolite LHVC has been subdivided into 3 main stages. The pre-caldera stage is represented by relatively abundant rhyolitic domes. Radiometric ages of some of these domes range between 693.0±1.9 ka ($^{40}Ar/^{39}Ar$, plagioclase) to 270±17 ka (U/Th, zircon) (Carrasco-Núñez et al., 2018) and 360±100 ka (sanidine) and 220±40 ka (sanidine) (K/Ar, whole rock, Ferriz and Mahood, 1984).

The caldera stage consists of two major caldera-forming events, separated by large Plinian eruptive phases (Fig. 1). The first and largest caldera-forming eruption, producing the trap-door Los Humeros caldera, is associated with the emplacement of the rhyolitic Xaltipan ignimbrite, with an estimated volume of 115 km$^3$ DRE (Ferriz and Mahood, 1984) and a $^{40}Ar/^{39}Ar$ radiometric age of 164.0±4.2 ka (Carrasco-Nuñez et al., 2018). Following this catastrophic event, a sequence of explosive episodes occurred at 70±23 ka ($^{40}Ar/^{39}Ar$, Carrasco-Núñez et al., 2018). These eruptions emplaced thick rhyodacitic Plinian deposits grouped into the Faby Tuff unit (Ferriz and Mahood 1984; Willcox, 2011). The second caldera-forming episode, very close in time with the Faby Tuff unit, produced the Los Potreros caldera, which is associated with the emplacement of the rhyodacitic to andesitic Zaragoza ignimbrite, with an estimated volume of 15 km$^3$ DRE (Carrasco-Núñez and Branney, 2005; Carrasco-Núñez et al., 2012) and a radiometric age of 69±16 ka ($^{40}Ar/^{39}Ar$, Carrasco-Núñez et al., 2018).

The post-caldera stage has been divided into two different intra-caldera eruptive phases (Carrasco-Núñez et al., 2018). The first one was a late Pleistocene phase characterized by the emplacement of rhyolitic and dacitic domes at about 50.7±4.4 - 44.8±1.7 ka ($^{40}Ar/^{39}Ar$ ages), followed by a sequence of explosive eruptions producing dacitic pumice fall units, volcaniclastic breccia and pyroclastic flows deposits with a maximum age of 28.3±1.1 ka (Rojas-Ortega, 2016; Carrasco-Núñez et al., 2018). The second one is characterized by alternated episodes of effusive and explosive volcanism with a wide range of compositions involving basaltic-andesitic and basaltic trachytic, trachyandesitic lava flows and dacitic, trachydacitic, andesitic and basaltic pumice and scoria fall deposits emitted by numerous monogenetic eruptive centers located in the LHVC (Ferriz and Mahood, 1984; Dávila-Harris and Carrasco-Núñez, 2014; Norini et al., 2015; Carrasco et al., 2017a, 2017b). This effusive activity was firstly considered within a range of 40-20 ka (Ferriz & Mahood, 1984), however, recent dating reveals that most of this activity is Holocene (Davila-Harris and Carrasco-Núñez, 2014; Carrasco et al., 2017b).

The LHVC hosts a geothermal field currently in exploitation, with an installed capacity of *ca*. 95 MW of electric power (Gutierrez-Negrín, 2019). Geothermal fluids are hosted by the 10.5-1.55 Ma andesites and basalts of the Cuyoaco, Alseseca and Teziutlán units, sealed upwards by the low-permeability Quaternary ignimbrites that act as cap rocks (Cedillo-Rodríguez, 1997, 1999; Arellano et al., 2003; Lorenzo-Pulido, 2008; Gutierrez-Negrín and Izquierdo-Montalvo, 2010), although highly variable welding *facies* are observed on the surface, which suggest a wide range of permeable conditions for the pyroclastic rocks. Conceptual models of the geothermal field are based on hydrothermal fluid circulation occurring along fault and fracture systems inside the Los Potreros Caldera (Cedillo-Rodríguez, 1997, 1999; Arellano et al., 2003). These faults have been interpreted as regional, partly sealed, tectonic structures (e.g. Cedillo-Rodríguez, 1997, 1999) or, more recently, as generated by the active/recent resurgence of the caldera floor (Norini et al., 2015).

### 3. Structural analysis of the LHVC pre-volcanic basement

The LHVC pre-volcanic basement, which consists of Precambrian–Paleozoic crystalline rocks, Jurassic and Cretaceous sedimentary rocks and Eocene–Quaternary intrusive and effusive magmatic rocks, has been extensively surveyed at the outcrop scale (Figs. 1 and 2). The attitude of the sedimentary rocks has been measured in more than 200 outcrops, to define the general setting of the basement correlated with the MFTB orogenic phase (Fig. 2). Also, geometric measurements and kinematic interpretation have been carried out on all the tectonic and intrusive structures identified in the sedimentary and magmatic rocks, to define the structural evolution of the area, the regional stress field, and their influence on the evolution of the caldera complex (Fig. 3).

The sedimentary basement surrounding and underlying the LHVC is pervasively folded and thrust faulted during the MFTB orogenic phase (e.g. Fitz–Diaz et al., 2017). The local structural style is defined by fractured thinly–bedded carbonates, interbedded with cherts and shales, affected by tight outcrop-scale *chevron* folds detached above km–scale thrusts with a dominant NE-directed tectonic transport (Figs. 1 and 3a,b). Field measurement of bedding in carbonates and terrigenous rocks shows an average attitude toward the SW and cylindrical folding at different wavelengths with NW–SE trending β axis (Fig. 2). The average dip of the sedimentary rocks gradually changes from 60°-65° in the SW sector, to 20°-40° in the central sector, to nearly horizontal bedding in the NE sector of the study area (see Figure 1). The decreasing average dip of the sedimentary pre-volcanic basement from the SW to the NE depicts a major ten–of–km–scale asymmetric anticline fold (Figs 1 and 2). North of the LHVC, this anticline fold exposes the crystalline metamorphic basement (Teziutlan Massif) in its core (Fig. 2) (e.g. Yáñez and García, 1980; Angeles–Moreno, 2012). Most of the internal deformation of the sedimentary rocks has been accommodated by the intense folding. Only few intra–formational thrust faults are clearly exposed in the area. These moderate– to low–angle faults (dip ≤ 50°) are associated with well–developed drag folds in the limestone, have NE vergence, and show slickenlines indicative of dip–slip reverse displacements (Fig. 3a,b, Table 1). The kinematic indicators measured on the thrust faults have been analyzed with the Right Dihedron method of Angelier and Mechler (1977). This analysis suggests that the area deformed in the Late Cretaceous–Eocene under a compressive stress field with NE–SW trending maximum horizontal stress (horizontal $\sigma_1$ and vertical $\sigma_3$) (Fig. 3b).

Few NE–striking normal faults have also been identified in the study area (Fig. 1). These discontinuous faults postdated the MFTB orogenic phase, displace the sedimentary basement and some Miocene magmatic intrusions (≈15 Ma) and are mostly sealed by volcanic units of the TMVB (<8.9-11 Ma) (Carrasco–Núñez et al., 2017b). The measured faults have dip angle ranging 60°–80° and slikenlines on the fault planes consistent with normal and transtensional faulting. The kinematic analysis suggests that this Eocene–Pliocene deformation occurred under an extensional stress field with NE–SW trending maximum horizontal stress (vertical $\sigma_1$ and horizontal $\sigma_3$) (Fig. 3c, Table 1).

The results of the kinematic analysis of the MFTB compressive structures and younger normal faults indicates that the direction of the maximum horizontal stress (first $\sigma_1$ and then $\sigma_2$), exerting a control on the direction of magmatic intrusions and transport of fluids in the crust, maintained a constant NE–SW trend during time (Fig. 3).

Small mafic dikes/sills and large granite-granodiorite magmatic intrusions of Miocene age emplaced in the pre-volcanic sedimentary rocks cropping out in the area surrounding the LHVC (Carrasco–Núñez et al., 2017) (Figs. 1 and 3d,e, Table 1). The geological map and geomorphological analysis of a 20 m DEM show that the large silicic magmatic intrusions, exposed at the surface as high-relief elongated ridges, have a NE–SW trend, parallel to the maximum horizontal stress of the two main tectonic phases identified in the study area (SGM, 2010a, 2011a, 2011b; Carrasco–Núñez et al., 2017b) (Figs. 1 and 3). The preferential orientation of Miocene large intrusions is also outlined by NE-SW-trending elongated sharp magnetic anomalies recorded in the area surrounding the LHVC (e.g. SGM, 2010b, 2014). Small Miocene-Pliocene mafic intrusions measured in the field show two preferential strikes (Fig. 3d,e,f, Table 1). In particular, most dikes emplaced with a NE–SW strike, parallel to the few scattered NE–striking normal faults and the maximum horizontal stress (Fig. 3d, Table 1), while sub-vertical sills emplaced parallel to the NW–SE striking sedimentary bedding tilted by tight folds, along the weak interfaces among the strata of the fractured carbonates (Fig. 3e, Table 1). An outcrop to the west of LHVC exposes a NE–SW-striking dike turning upward into a sill parallel to the NW–SE-striking sedimentary bedding (LH2017-102 in Table 1). The preferential trends of mafic and silicic intrusions suggest that magmatic fluids in the crust have moved along two orthogonal preferential directions, NE–SW and NW–SE, controlled by the regional stress field and the inherited weakness planes in the pre-volcanic basement (Fig. 3f, Table 1).

| Outcrop | Latitude | Longitude | Elevation (m a.s.l.) |
| --- | --- | --- | --- |
| PDL52 | 97° 26' 35.599" W | 19° 40' 41.173" N | 2820 |
| PDL54 | 97° 27' 20.991" W | 19° 40' 28.698" N | 2802 |
| LH2017_36 | 97° 27' 17.363" W | 19° 41' 11.252" N | 2819 |
| LH2017_40 | 97° 26' 46.658" W | 19° 40' 21.616" N | 2820 |
| LH2017_43 | 97° 27' 43.91"" W | 19° 40' 59.704" N | 2764 |
| LH2017_101 | 97° 43' 59.301" W | 19° 36' 18.101" N | 2398 |
| LH2017_102 | 97° 35' 59.424" W | 19° 41' 35.840" N | 2558 |
| LH2017_108 | 97° 29' 37.417" W | 19° 29' 36.007" N | 2376 |
| LH2017_115 | 97° 12' 19.714" W | 19° 50' 49.100" N | 1452 |
| LH2018_02 | 97° 24' 12.764" W | 19° 23' 35.974" N | 2359 |

Table 1: location of outcrops described in the text and showed in figures 2, 3, 4 and 5.

## 4. Volcanotectonic deformations of LHVC: geomorphology and field data

The geomorphology of the Los Humeros and Los Potreros calderas have been analyzed to identify and map lineaments and fault scarps. These surface deformation features were interpreted from shaded relief images processed from a new and accurate 1 m resolution DEM covering an area of 300 km$^2$. The DEM has been computed by Rational Polynomial Coefficient (RPC) digital photogrammetry of a GeoEye1 satellite panchromatic (450–800 nm) stereo pair with 0.5 m resolution. The high–resolution DEM allowed to redefine and reinterpret with increased precision the volcanotectonic structures discussed by Norini et al. (2015). All these structures have also been verified in the field, where geometric measurements and kinematic interpretations have been conducted on the observed deformation features. Unfortunately, the area is widely forested and most of the volcanic products exposed in the few outcrops are unconsolidated pyroclastic deposits and/or fractured lava flows and domes with very poor preservation of the kinematic indicators on the fault planes. Nevertheless, the geometric and kinematic study of some of the major faults has been made possible by excavations conducted in recent years in the frame of the CFE's study of the geothermal concession.

The high–resolution DEM shows semicircular topographic reliefs corresponding to sharp morphological scarps and elongated ridges covered by post–caldera volcanic units, which have been used to identify and interpret the caldera rims (Figs. 1 and 4a) (Carrasco–Núñez et al., 2017). These caldera rims have semicircular asymmetric plan view shapes, which are consistent with the inferred trap–door collapse structure of the Los Humeros collapse caldera (e.g. Norini et al., 2015; Carrasco–Núñez et al., 2017) (Fig. 1). The Los Humeros collapse caldera has a maximum diameter of 17–18 km (Fig. 1). The hinge of this trap–door caldera, where the morphological rim vanishes and has no topographic relief, is located to the NE, while the maximum vertical displacements along the inferred ring faults occurred in the southwestern sector of the caldera border (Fig. 1). The Los Potreros collapse caldera is located inside the larger Los Humeros caldera, and has a diameter of 9 km and a nearly–mirror pattern of the rim if compared with the Los Humeros caldera. The semicircular morphological rim of the Los Potreros caldera is clearly visible toward the E–NE and disappears in the opposite direction (Fig. 1 and 4a).

The orthorectified 0.5 m GeoEye1 image, high–resolution DEM and detailed fieldwork improved the mapping of the ≈100 monogenetic volcanic centers emplaced within the caldera complex after the Los Potreros caldera collapse (<69 ka) (e.g. Norini et al., 2015; Carrasco–Núñez et al., 2017b). The spatial distribution of these pyroclastic cones and eruptive fissures depicts a NNW-SSE elongated ring-shaped area parallel to main MFTB faults and fold axis (Norini et al., 2015) (Fig. 1).

The structural analysis focused mainly on a distinct fault system that has been previously related to the active resurgence of the LHVC (e.g. Norini et al., 2015). The fault system crops out inside the ring-shaped area defined by the spatial distribution of post-caldera monogenetic eruptive centers (Fig. 1) (Norini et al., 2015). The topographic expression of the fault system is represented by rectilinear and curvilinear prominent fault scarps with a maximum elevation of 70–80 m above the caldera floor (Fig. 4). These active/recent faults traverse the geothermal area under exploitation and are correlated with the enhanced secondary permeability of the hydrothermal system (Norini et al., 2015). The main fault swarm, represented by the Maxtaloya fault, the Los Humeros fault and some sub-parallel fault strands, is slightly curvilinear and runs NNW–SSE for ≈8 km (Fig. 4a). This fault swarm is characterized by strong hydrothermal alteration of the displaced volcanic units and delimits to the west the deformed area and the relief generated by the Los Potreros caldera rim (Figs. 1 and 4a). Kinematic data measured in a man-made outcrop along the Los Humeros fault scarp (PDL54 in Figs. 4a and 5a, Table 1) indicate dominant dip-slip normal displacements of pumice fall deposits and lava flows, with minor changes of the displacement vector related to the bending of the fault trace (Fig. 5a). The high–resolution DEM shows multiple N–S, NE–SW and E–W curvilinear splays departing from the main NNW–SSE fault swarm and depicting a complex deformation pattern (Fig. 4a). Both the main fault swarm and its splays vanish when approaching the Los Potreros caldera rim, with a marked decrease of the fault scarps height toward the periphery of the fault system (Fig. 4). Also, all these structures invariably displace a marker pyroclastic fall deposit with $^{14}$C radiometric age of 7.3±0.1 ka, indicating volcanotectonic deformation along Holocene faults (Dávila-Harris and Carrasco–Núñez, 2014; Norini et al., 2015) (Fig. 5a,b,c).

As previously proposed by Norini et al. (2015), two distinct resurgent structural sectors occur within the half–moon shaped area delimited by the main NNW–SSE faults swarm to the west and the Los Potreros caldera rim to the east and north (Fig. 4a). The southern resurgence sector (S1 in Norini et al., 2015) is bounded to the west by the NNW–SSE Maxtaloya fault and to the north by the NE–striking Arroyo Grande fault (Fig. 4). The Maxtaloya fault shows mainly normal and minor right-lateral displacements of marker features in post-caldera < 69 ka lava flows and pyroclastic fall deposits (Carrasco–Núñez et al., 2017b) (Fig. 4a). The NE–striking Arroyo Grande fault has been exposed by excavation along a small sector of the fault scarp (PDL52 in Figs. 4a and 5c, Table 1). At the base of this man-made outcrop, the main fault plane affects strongly hydrothermally altered pumice fall deposit and exposes well–preserved dip–slip slickenlines indicating reverse displacements under a compressive stress (Fig. 5c). In the upper part of the same outcrop, normal faults occur as secondary features accommodating the gravitational instability generated by the 80 m high fault scarp. The geometry at depth of the Arroyo Grande fault and other faults in the area may be constrained by geophysical and well logs data (e.g. *Section 5*). At its southern tip, the Arroyo Grande Fault and another parallel fault strand expose an uplifted 50-60 m thick section of the Zaragoza Ignimbrite in intra-caldera welded massive *facies*, interpreted as an evidence of volcanotectonic uplift/resurgence of the caldera floor (LH2017-40 in Fig. 4a, Table 1). The interior of the southern structural sector is characterized by the absence of visible hydrothermal alteration in the exposed pyroclastic units and is affected by several E–W/WNW–ESE striking, mainly dip–slip, sub–vertical faults, interrupting a gentle SSE–dipping monocline (Fig. 4a,b). The longest of these structures is the Las Papas fault, exposing fresh, non-altered Holocene pyroclastic fall deposits (Fig. 4).

The northern resurgent sector (S2 in Norini et al., 2015) is delimited to the east by the Arroyo Grande reverse fault and to the west by NNW-SSE fault strands (Fig. 4a,b). The westernmost structure bounding the northern sector displaces a fresh Holocene pahoehoe lava flow (Carrasco–Núñez et al., 2017b) and generates a sharp rectilinear 10 m high topographic scarp, clearly visible in the field and high–resolution DEM (LH2017-43 in Figs. 4a,b and 5d, Table 1). The brittle rheology of the young fractured lava flow prevented the exposure of fault planes and kinematic indicators. This fault scarp is located near the H43 geothermal well (see Fig. 4a for location), where reverse faults with a geometry compatible with the outcropping structure have been identified at depth (*Section 5.2*). To the east of the fault scarp, a series of N-S/NNE-SSW striking normal faults delimiting narrow grabens displace the doming topographic surface of the resurgent sector (Fig. 4a,b). Widespread hydrothermal alteration of pyroclastic deposits and lavas occurs along the fault scarps of this sector. Kinematic data measured on fault planes in a CFE's man-made outcrop indicate normal faulting kinematics with a minor component of left-lateral motion displacing a hydrothermally altered pumice fall deposit (LH2017-36 in Figs. 4a and 5b, Table 1). Imaging of this outcrop with a hand-held thermal camera showed sharp anomalies originated by hot fluid flow along the fault planes (Fig. 5b).

## 5. Well logs and geophysical data

The volcanotectonic analysis of Los Humeros and Los Potreros calderas allowed the identification of recent/active faults with complex and discontinuous surface traces. The analysis of the geometry of these faults and collapse structures requires subsurface data to unravel the structural geometry of the LHVC interior. To gain insights into the geometry, attitude, and possible intersections at depth of the main volcanotectonic structural features and the subsurface geology of the LHVC, a small set of well logs and seismological and geophysical data has been integrated in a 3D GIS database, along with the structural data collected in the field.

### *5.1. Lithological well logs*

More than 60 geothermal wells have been drilled within and around the Los Potreros caldera since the 1980 (e.g. Cedillo–Rodríguez, 1999; Arellano et al., 2003; Gutiérrez–Negrín and Izquierdo–Montalvo, 2010; Rocha López et al., 2010; Carrasco–Núñez et al., 2017). The lithological logs of some of these wells have been reclassified following the UBSUs stratigraphic criterion proposed by Norini et al. (2015) (Fig. 6). Units *Calderas phase* and *Post-calderas volcanism* of Norini et al. (2015) have been grouped into a single unit because of the scale of the work (unit *LHVC* in Fig. 6b), and, thus, the lithological logs of the geothermal wells have been reclassified into three UBSUs corresponding to: (1) *Sedimentary Basement*, (2) *Old-volcanic succession*, and (3) *LHVC* units (Fig. 6b). These UBSUs have been defined by the unconformities *Un1*, a regional non-conformity marking a temporal hiatus between the deposition of the sedimentary basement and the onset of the volcanic activity in the study area, and *Un2*, an evident angular unconformity associated with a temporal hiatus and a shifting of the feeding system, also marking a change in eruptive style, from the mainly effusive activity of the Teziutlán volcanic unit to the more silicic explosive/effusive products of the LHVC (Fig. 6b and Norini et al., 2015). The reference stratigraphic framework also includes a basal UBSU corresponding to the rocks below the regional non-conformity *Un0*, marking the temporal hiatus between the emplacement of the Precambrian–Paleozoic crystalline rocks (unit *Teziutlán Massif* in Fig. 6b) and the deposition of the Mesozoic–Paleogene marine sedimentary succession (unit *Sedimentary Basement* in Fig. 6b).

The lithological logs that have been included in our work are from the geothermal wells H5, H8, H10, H14, H17, H18, H19, H20, H21, H23, H24, H25, H27, H28 and H32. The true depth (TD) of the unconformities recognized in these wells, the stratigraphic contacts identified in the map of Fig. 1 and the attitude of geological structures measured in the field have been used to analyze the geometry of the caldera complex and basement and to draw two schematic geological cross-sections (Fig. 6c,d). TDs of the unconformities observed in the wells located near each section have been projected on the section plane. Also, some wells have been considered because of their relationship with the caldera morphological rims and faults, like wells H14 and H18 showing the change in elevation of the sedimentary basement across the Los Humeros ring fault (Fig. 6). The structures depicted in the two geological cross-sections within LHVC have also been interpreted form data discussed below (*Sections 5.2, 5.3 and 5.4*) and from Norini et al. (2015).

### *5.2. Formation Micro Imaging log of H43 geothermal well*
A Formation Micro Imaging (FMI) log has been recorded in the vertical H43 well in the north-western sector of the Los Potreros caldera (H43 in Fig. 4a). This electrical resistivity log provided microresistivity images of the borehole wall, useful for the identification and measurement of structural features in the encountered geological formations. Two segments of the well have been imaged by FMI: the first (1250-1634 m TD) is in Miocene-Pleistocene lavas and the second (1711-1813 m TD) is in Mesozoic limestone (Lorenzo-Pulido, 2008; Rocha-López et al., 2010). The FMI recorded the attitude of faults and fractures in the volcanic and sedimentary rocks. These brittle structural features are related to the secondary permeability in the volume of rocks hosting the geothermal system (Rocha-López et al., 2010). Their geometry is also important to depict a comprehensive structural view of the caldera complex, from the surface (faults and fractures measured in the field, *Section 4*) to the subsurface (FMI log). For this reason, we focused our geometric analysis and kinematic interpretation of the H43 FMI log on faults and Induced Fractures (IFs). IFs develop during drilling when the wellbore stress concentration is greater than the tensile strength of the rock. The minimum circumferential stress around a vertical borehole occurs parallel to the maximum horizontal stress and therefore IFs are open parallel to this principal stress and invaded by drilling mud (e.g. Tingay et al. 2005).

All faults imaged by FMI have a constant N-S/NNE-SSW strike, parallel to faults observed in the field in the same sector of the Los Potreros caldera (Figs. 4a and 7a). Faults in H43 dip eastward (dip angle ≈50-60°) (Rocha-López et al., 2010). Direction of IFs, expected to be parallel to the current maximum horizontal stress, shows a sudden change with depth along the well. IFs have NNE-SSW strike from the topographic surface to 1500 m TD, parallel or slightly oblique to the faults observed in the FMI log (Fig. 7b). This configuration, with IFs and maximum horizontal stress roughly parallel to faults, is compatible with normal faulting and extensional stress regime in the upper half of the geothermal field. At greater depth, IFs and the maximum horizontal stress have WNW-ESE strike (below 1500 m TD), roughly perpendicular to the observed faults, which is a configuration compatible with reverse faulting and compressive stress regime in the lower half of the geothermal field (Fig. 7c,d).

### *5.3. Focal mechanism of 16/08/2015 and 08/02/2016 earthquakes*

A small broad band seismic network has been deployed in the Los Humeros geothermal field since the 90's (Lermo et al. 2008, 2016; Antayhua et al., 2008) (Fig. 8a, inset). The network recorded several induced and natural earthquakes of small magnitude (typically Md < 2) within the geothermal field (e.g. Lermo et al. 2008). To obtain the hypocentral parameters of the seismic events a local velocity model has been calculated by Lermo et al. (2001), resulting in localization errors of less than 0.3 km (latitude, longitude and depth). Few stronger earthquakes have also been recorded by the seismic network, and for some of them focal mechanism solutions have been calculated (Lermo et al., 2016).

Two of the most recent and significant earthquakes in the geothermal field occurred on August 16, 2015 at 14:29 and on February 08, 2016 at 21:16 GMT (Table 2) (Lermo et al., 2016). The epicentres of these events are located along the trace of the Maxtaloya-Los Humeros fault swarm (Fig. 4). The focal mechanism solutions of the two earthquakes have been recalculated from the first arrival polarity of the vertical component recorded by the seismic stations and incorporated in the 3D GIS database to analyse their relationships with other surface and subsurface data. Both focal mechanisms indicate reverse faulting along most reliable planes of rupture, defined by aftershocks (Lermo et al., 2016), striking NNW-SSE (Fig. 8 and Table 2). The geometry of the rupture planes is compatible with some of the measured structures identified in the vicinity of the epicenters (Figs. 4a, 5, 8a) and to the faults imaged by FMI in well H43 (Fig. 7a). The reverse sense of slip also agrees well with some observations made in the field and FMI log (Figs. 4a, 5c and 7d), suggesting ongoing reverse faulting and compressive stress regime in the lower part of the geothermal field (hypocentres at 1.6 and 1.9 km depth) (Fig. 8b) (Table 2).

| Date | Time (GMT) | Latitude (°) | Longitude (°) | Depth (km) | Mw | Rupture plane | | |
|---|---|---|---|---|---|---|---|---|
| | | | | | | Strike (°) | Dip (°) | Rake (°) |
| 16/08/2015 | 14h 29m 16.5s | 19.68952 | -97.452 | 1.6 | 2 | 332 | 61 | 42 |
| 08/02/2016 | 21h 16m 2.7s | 19.66822 | -97.454 | 1.9 | 4.2 | 169 | 61 | 42 |

Table 2: time, location and source parameters of the 16/08/2015 and 08/02/2016 earthquakes.

### *5.4. Magnetotelluric resistivity imaging of the caldera complex*

MT resistivity imaging of the geothermal field has been recently obtained from the surface up to a depth of 15 km (Arzate et al., 2018). From the original 70 MT soundings, acquiring time series during periods ranging from 20 to 24 hours at each site using four Phoenix Geophysics magnetotelluric devices, a new NW-SE profile has been extracted and processed (Figs. 1 and 9a). This new profile passes through 14 MT stations and has been focused on the Maxtaloya-Los Humeros fault swarm and the Los Humeros caldera SW rim using a high resolution frequency range (from $10^4$ to $10^{-1}$ Hz), roughly corresponding to a penetration depth of about 3 km for an average ground resistivity of 3 ohm-m (Vozoff, 1972). During the acquisition, the four available devices were synchronously running in order to estimate the impedance tensor using the remote reference processing and improving the apparent resistivity and phase curves (Gamble et al., 1979). At each station, two components of the electrical field (eNS, eEW) and three of the magnetic field (hNS, hEW, hz) have been measured. The acquired time series were processed using standard fast Fourier transform algorithms and robust cascade decimation (e.g. Wight and Bostick, 1980; Simpson and Bahr, 2005). The resulting impedances were then converted to resistivity and phase field curves that provided the basis for the electrical model. Determinant-based 2D modelling has been performed calculating the determinant of the impedance tensor and the 2D NLCG algorithm has been applied (Rodi and Mackie, 2001; Pedersen and Engels, 2005). The residuals from the normalized difference between the observed resistivity and phase curves and the fitted curves resulting from the 2D inversion for the most part of the spectra remain below 10%, which is indicative of a good correlation of the profile inversions.

The MT profile ranges from highly conductive (< 5 ohm-m) to resistive values (≈300-500 ohm-m), with sharp resistivity changes within the geothermal field, mainly related to lithology, hydrothermal alteration and/or occurrence of hydrothermal fluids (Fig. 9a) (Arzate et al., 2018). From southwest to northeast most of the changes in resistivity correspond to structures already identified in the geological map and cross-sections (Figs. 1 and 9). In the SW, the MT profile shows a sharp resistivity gradient spatially corresponding to the inferred SW boundary of the Los Humeros caldera (Figs. 1 and 9). This resistivity contrast dips outward (south-westward) respect to the caldera centre, in agreement with the inferred location and geometry of the caldera ring fault (e.g. Norini et al., 2015) (Fig. 9a,b). North-eastward, two marked changes in resistivity define inward (north-eastward) dipping curved trends (Fig. 9a). These resistivity contrasts spatially correspond to caldera resurgence structures identified in the field, geological map, geological cross-sections and the 08/02/2016 focal mechanism solution (Figs. 1, 6, 9b,c). Also, faults with same attitude and location have been imaged to the north by the H43 FMI log (Fig. 7d). In the central portion of the MT profile, a shallow resistive sector (blue in Fig. 9a,b) is delimited by a more conductive concave zone (green in Fig. 9a,b). This resistive shallow body corresponds in the field to the caldera resurgence area where hydrothermal alteration is absent in the exposed pyroclastic deposits (e.g. Las Papas fault, Figure 4). North-eastward, the MT profile imaged a south-westward shallow dipping resistivity contrast roughly corresponding to the lithological contact between the sedimentary basement and the overlying volcanic succession (Fig. 9). The attitude of these gently dipping contact supports the inferred trap-door collapse geometry of the Los Humeros caldera (e.g. Norini et al., 2015).

## 6. Discussions
### *6.1. Volcanotectonic interplay in the LHVC*

Interplay among geological processes at different spatial and temporal scales occurs in the LHVC area. This interaction generated a complex structural grain, where both inherited and active tectonic/volcanotectonic structures have contributed to the evolution of the caldera complex. The combination of morphostructural analysis, field data and 3D analysis of well logs, together with geophysical imaging and seismological data show that LHVC deformation style resulted from the coexistence of the following regional and local structural elements (Figs. 6c,d and 10a):

a) Late-Cretaceous-Eocene MFTB orogeny, with a roughly SW–NE trending regional $\sigma_{hmax}$, generated pervasive folding and thrusting in the crystalline and sedimentary basement underlying the TMVB sequence (Figs. 1, 2). The general strike of folds and thrust faults is NW-SE/NNW-SSE (Fig. 10a). The most important inherited regional structure within the study area is a wide km-scale anticline fold, exposing the crystalline basement in its core to the north of the caldera complex. The NNW-SSE-trending fold axis passes beneath LHVC (Fig 10a). Even if these orogenic structural features are inactive, they have exerted a clear control on the rising and emplacement of TMVB magmas. Field observations show that intrusion of Miocene-recent mafic sills occurred along sub-vertical and vertical weak planes corresponding to bedding of the sedimentary basement tilted by tight folds (Fig. 3e,f). Low tensile strength of these bedding planes allowed propagation of hydraulic fractures driven by magma excess pressure (e.g. Gudmundsson, 2011);

b) A weak extensional tectonic phase occurred in the LHVC area since the Miocene (Fig. 3c). The SW–NE trending regional $\sigma_{hmax}$ was responsible for the brittle deformation of the crust along steep SW-NE-striking normal faults and extensional fractures (Fig. 10a). This tectonic phase occurred during the emplacement of TMVB magmas, controlling the geometry of large SW-NE-trending silicic intrusions and smaller mafic dikes, parallel to the faults, fractures and regional $\sigma_{hmax}$ (Fig. 3d,f). Most extensional structures appear to be older and sealed by deposits related to LHVC activity (Fig. 1) (Carrasco–Núñez et al., 2017b), even if they played a role in the structural architecture of the caldera complex (see points d and f);

c) Few data and constraints are available on the geometry of the LHVC feeding system. The caldera complex activity started since Middle-Late Pleistocene (164 ka), with the emplacement of a large magma chamber at about 5 to 7-8 km depth in the crust (Martínez et al., 1983; Verma, 1983, 2000; Campos-Enríquez and Garduño-Monroy, 1987). Considering the geometry of the main tectonic structures proposed in this work, the location of LHVC roughly corresponds to the axis of the MFTB major anticline fold identified in the sedimentary and crystalline basement (Fig. 10a). LHVC magma conduits cross the fold at depth. This configuration suggests the possibility that the unfavorable geometry of the anticline fold, with horizontal/low-angle weak planes in the basement, may have represented a trap for rising magma, promoting hydrofractures arrest, sill formation and the emplacement of the magma chamber (e.g. Gudmundsson and Brenner, 2001; Gudmundsson, 2011; Ferré et al., 2012; Norini et al., 2013);

d) The collapse of the trap-door Los Humeros caldera occurred along sub-vertical outward-dipping ring faults (Figs. 6c,d and 9) (Norini et al., 2015; Carrasco–Núñez et al., 2017b). In its southeastern sector, the caldera morphological rim is rectilinear and roughly parallel to the NE-SW-striking TMVB Miocene-recent normal faults, extensional fractures and regional $\sigma_{hmax}$ (Fig. 10a). This geometry suggests that caldera ring faults may have partly reactivated inherited steep tectonic structures in the basement as weak planes for the collapse of the magma chamber roof (Figs. 6d and 10a);

e) Several monogenetic volcanic centers have been emplaced within the caldera complex after the Zaragoza ignimbrite eruption (Carrasco–Núñez et al., 2017b). The spatial distribution of these volcanoes defines a NNW-SSE elongated ring-shaped structure where density of monogenetic centers is higher than in the surroundings (Figs. 1 and 10a) (Norini et al., 2015). The orientation of the ring-shaped volcanotectonic feature indicates that the activity of monogenetic volcanoes is fed by a magmatic intrusion elongated in the same NNW-SSE direction. This geometry is parallel to the main MFTB structures near LHVC, suggesting that the magma chamber of the caldera complex is drained by NNW-SSE hydrofractures (e.g. Norini et al., 2008, 2013, 2015). The trend of the elongated ring-shaped feature roughly follows the strike of MFTB main anticline fold and bedding weak planes along which mafic sills intruded the sedimentary basement (Fig. 3e,f). Thus, the emplacement of post-calderas LHVC magmas feeding monogenetic volcanoes may have been controlled by inherited tectonic structures promoting hydrofractures propagation and arrest in the basement;

f) in the post-caldera phase, resurgence of Los Potreros caldera floor induced local deformations in the crust (Norini et al., 2015). Faults with different geometry and kinematic displace by tens of meters Upper-Pleistocene-Holocene volcanic deposits in the center of the caldera, showing a marked decrease in fault scarps height and displacements toward the caldera rims (e.g. Fig. 4a,b). Norini et al. (2015) calculated a maximum vertical displacement rate of 10 mm/yr along these faults during the Holocene. Reverse and normal displacements along faults exposed in the field are the expression of changing stress field in the subsurface, as documented by well logs and geophysical data (Figs. 4, 5, 7, 8 and 9, *Section 6.2*). The geometry and kinematic of recent-active fault strands in the Los Potreros caldera floor depict a local radial stress field, possibly related to magmatic intrusions and/or pressurization of the hydrothermal system (Fig. 10a) (Norini et al., 2015). Also, minor components of left-lateral (Los Humeros fault) and right lateral (Maxtaloya fault) motions along the mainly dip-slip NNW-SSE sub-vertical fault planes (Fig. 4a) are compatible with a trap-door uplift of the resurgent caldera floor, exposing in the center of the caldera an ignimbrite deposit that can be correlated to the Zaragoza Ignimbrite in the form of intra-caldera welded massive *facies* (outcrop LH2017-40 in Fig. 4a). The volcanotectonic deformation occurred simultaneously with post-caldera monogenetic eruptive activity (Fig 10a). Both post-caldera monogenetic volcanoes and volcanotectonic deformation may be associated with the emplacement of small magmatic intrusions below the Los Potreros caldera (Norini et al., 2015). Resurgence faults and structure of the Los Humeros geothermal field are discussed in detail in *Section 6.2*.

The points mentioned above indicate that during the evolution of LHVC the main mechanism of volcanotectonic interaction has been the use of inherited weak fault and bedding planes for: (1) reactivation of favorable pre-existing mechanical discontinuities to accommodate displacements along ring faults and resurgence faults, and (2) magma rising and emplacement through hydrofractures propagation and arrest along low tensile strength planes. These processes occurred in a lithological and mechanical heterogeneous crust (metamorphic, sedimentary and magmatic overlapping units, Figs. 1 and 6) under a weak extensional stress field with SW–NE trending regional $\sigma_{hmax}$ (Fig. 10a). Inside the caldera complex, the local radial stress field of magmatic origin overwhelms the far-field tectonic stress, inducing the formation of faults and fractures in the Los Humeros geothermal reservoir.

### *6.2. Implications for the structure of Los Humeros geothermal reservoir*

Geothermal manifestations and fracture-controlled secondary permeability in the Los Humeros geothermal field are linked with the complex volcanotectonic fault system displacing the Los Potreros caldera floor (*Section 4*, Fig. 4a and thermal image in Fig. 5b) (Norini et al., 2015). This fault system defines two resurgent structural sectors delimited by the NNW-SSE Maxtaloya-Los Humeros fault swarm, the Arroyo Grande fault and the Los Potreros caldera rim (Fig. 4a). The NNW-SSE Maxtaloya-Los Humeros fault swarm is parallel to the MFTB inherited structures, while the NE-SW Arroyo Grande fault and parallel fault strands have the same strike of the TMVB normal faults and regional $\sigma_{hmax}$ recognized in the area (Fig. 10b). As already discussed in *Section 6.1*, this geometric link suggests the possibility that the main volcanotectonic features in the Los Potreros caldera floor have been formed by reactivation of inherited weak planes generated by regional tectonics in the LHVC basement.

The surface trace of the volcanotectonic fault system is well constrained by high-resolution geomorphological and field data (*Section 4*). At depth, the 3D geometry of the main faults has been constrained by a small set of well logs and seismological and geophysical subsurface data (*Sections 5.2*, *5.3* and *5.4*). The H43 FMI log, 08/02/2016 earthquake focal mechanism solution and SW-NE MT resistivity profile are all independent lines of evidence suggesting inward dipping geometry respect to the caldera centre of NNW-SSE and NE-SW fault strands (Figs. 7d, 8b and 9). These faults define listric steep ramps down to 3 km depth or more below the topographic surface (Figs. 6c,d and 9b,c). The 3D geometry of these fault ramps may be better constrained in the future by analysis of the pressure and temperature profiles in the geothermal wells. The Arroyo Grande fault represents one of these ramps reaching the topographic surface, where reverse displacements of pyroclastic deposits have been measured in the field (Figs. 4a and 5c). Other fault ramps may be blind under the thick cover of post-caldera volcanic products and some of them have been identified by subsurface data (Figs. 6c,d, 7d and 9b,c). Conjugate, westward and northwestward dipping normal faults originate from the main fault ramps at shallow depths (Figs. 5a, 6c and 9c).

In outcrop, faults with long surface trace (> 1.5 km), expected to reach depths where hydrothermal fluids circulate in the reservoir (e.g. Cedillo, 1999; Arellano et al., 2003), invariably show strong hydrothermal alteration of the displaced volcanic units, except for the E-W striking Las Papas fault in the southern resurgent sector (*Section 4* and Fig. 10b). The occurrence of hydrothermal alterations and geothermal manifestations along volcanotectonic faults delimiting resurgent sectors suggests that the geothermal fluids are driven directly to shallow levels along these fault ramps, whose strikes have been inherited from the basement tectonic structures (Figs. 4a, 9b and 10b). In the southern resurgent sector, the E-W Las Papas fault, exposing fresh, non-altered pyroclastic fall deposits, corresponds to a shallow resistive sector in the MT profile (Fig. 9a). The lack of hydrothermal alterations and geothermal manifestations along the E-W structure, with very high resistive values in the MT profile, suggests that the Las Papas fault and parallel fault strands are shallow structures with absent or very weak connection with the geothermal reservoir. This configuration indicates that the southern structural sector behaves mainly as a single monolithic block uplifted by resurgence, with few shallow E-W faults accommodating minor internal deformations (Figs. 4a and 10c). The northern resurgent sector is affected by normal faults delimiting narrow N-S/NNE-SSW grabens (Fig. 4a). These structural features accommodate doming of the topographic surface, inducing uplift of the caldera floor and extension at shallow depths (Figs. 4b and 10c). At greater depth (> 1.5 km below the surface), subsurface data indicate reverse faulting and compressive deformation (Fig. 7c,d, 8 and 10c).

Doming/uplifted topography, inward dipping main fault ramps with respect to the caldera center and changes in stress field, from extension to compression, point to a common pressure source below the caldera floor as the origin of the local radial stress field and caldera resurgence (Fig. 10a,c). This local stress field, promoting radial fractures and faulting in the area, is compatible with the structural field data (*Section 4*) and subsurface data, including borehole imaging log and focal mechanism solutions (*Section 5*). The most probable origin of the local stress field is the pressurization of the shallow magmatic and/or hydrothermal system of the caldera complex, with a process similar to that described in Montanari et al. (2017) (Fig. 10c). Norini et al. (2015) proposed that the magmatic system feeding post-calderas monogenetic volcanic activity may be responsible for volcanotectonic deformations and caldera resurgence. Changes in pressure of the shallow magmatic/hydrothermal system may induce cyclic inversion of the dip-slip resurgence faults (cycling shifting from uplift to subsidence). Also, operation of the geothermal field, with extraction and injection of fluids in the crust, may have generated in the last 30-40 years ground subsidence that could overwhelm the long-term resurgence deformation signal (Békési et al., 2019).

In our structural model of LHVC, the complex 3D geometry of resurgence faults is the main volcanotectonic factor affecting distribution of secondary permeability within the geothermal reservoir. The radial stress field influences the strike of hydrofractures, with the expected geometry of faults and fractures producing geothermal fluids varying with location and depth (e.g. Figs. 7b,c and 10a,c).

**7. Final remarks**

The principal contributions of our study to the knowledge of the LHVC can be summarized by the following points:

1) our structural analysis suggests that at least two different orders of inherited regional tectonic structures played a role in the evolution of the magma feeding system, caldera collapses and post-caldera deformations of LHVC. These regional systems are the MFTB regional folds and inherited weak bedding/fault and fracture planes as well as the TMVB normal faults and extensional fractures. Both tectonic systems were generated under a regional NE-SW trending $\sigma_{hmax}$ and postdated the LHVC activity;

2) Local radial stress field, overwhelming the regional stress field and induced by the shallow LHVC magmatic/hydrothermal system, induces caldera resurgence and volcanotectonic faulting. Deformation of the caldera floor occurs simultaneously with post-calderas monogenetic activity distributed around the resurgence area;

3) Main resurgence faults and post-caldera magma-driven hydrofractures reactivated the inherited tectonic weak planes in the basement underlying the LHVC. Geometry and kinematics of resurgence faults are controlled by inherited regional structures and volcanotectonic radial stress field;

4) The permeability in the reservoir is mainly secondary and related to the damage zone of resurgence faults and, to a less degree, to inherited pervasive tectonic deformations. Complex geometry of resurgence faults and local volcanotectonic stress field are the main factors affecting the variability of secondary permeability within the hydrothermal system and the location of structures producing geothermal fluids.

This view of the caldera complex is fundamental to improve success rate in the exploitation of the shallow geothermal field (convective hydrothermal system at < 3 km depth), future exploration of deeper Super-Hot Geothermal Systems (SHGSs) near the magmatic heat source and engineering of Enhanced Geothermal Systems (EGSs). Understanding of volcanotectonic interplay in LHVC is of paramount importance, not only for the production of geothermal energy, but also because the availability of surface and subsurface data and the resulting 3D structural view make this volcano an important natural laboratory for the development of general models of volcano-tectonic interaction in calderas.

In our opinion, the next step to improve the structural knowledge of LHVC should be based on the analysis of pressure-temperature profiles of geothermal wells, perforation cuttings and detailed high-resolution geophysical data, as well as analogue and numerical modelling, to test the proposed structural model, assess the possible future structural evolution of the caldera complex, and evaluate the best strategies for geothermal exploration and production.

**Acknowledgments**

The research leading to these results has received funding from the GEMex Project, funded by the European Union's Horizon 2020 research and innovation programme under grant agreement No. 727550. This work also is a contribution to the project P05-CeMie-Geo No. 207032 SENER-CONACYT. The Comisión Federal de Electricidad of Mexico provided logs of geothermal wells and access to the geothermal concession area. The Centro de Geociencias of the Universidad Nacional Autónoma de México provided vehicles for the field campaigns. We acknowledge Gianluca Groppelli, Silvia Massaro and Roberto Sulpizio for their help in the fieldwork. We acknowledge Christian Ordaz, Anna Jentsch, Egbert Jolie, Guido Giordano, Federico Lucci, Federico Rossetti, Domenico Liotta, Andrea Brogi and the GEMex team for useful discussion, and the Comisión Federal de Electricidad personnel for their support during the study of the geothermal field. K. Bär and C. Rochelle kindly provided outcrop photograph of Fig. 3d and thermal image of Fig. 5b. We acknowledge Lucia Capra, Luca Ferrari and Alessandro Aiuppa for their valuable and constructive comments.

**Figure captions**

**Figure 1**: Simplified geological map of the Los Humeros Volcanic Complex (LHVC) and surrounding basement, on a shaded relief obtained from a 20 m resolution DEM (illuminated from the NW). The traces of the A-A' and B-B' geological cross-sections (Fig. 6c,d) and of the MT-MT' magnetotelluric profile (Fig. 9) are shown. In the inset, location of the LHVC in the frame of the Trans-Mexican Volcanic Belt (TMVB).

**Figure 2**: Schematic map of the main basement units exposed around LHVC and attitude of the Mesozoic sedimentary rocks, on a Landsat satellite image. Teziutlan Massif: Precambrian–Paleozoic crystalline basement, made of metamorphic and intrusive rocks, including green schists, granodiorites and granites. Sedimentary basement: Jurassic-Cretaceous limestone and terrigenous sedimentary rocks. The inset box shows the density of poles normal to bedding planes measured in the sedimentary basement (equal area projection, lower hemisphere). Location of outcrops in Fig. 3 is shown.

**Figure 3**: Photographs, stereographic projections and rose diagrams of fault and fracture data collected in the LHVC sedimentary and magmatic basement. Location in Fig. 2. (a) NW-SE-striking thrust fault of the sedimentary basement in outcrop LH2017_108. (b) Stereographic projection of the fault data measured in outcrop LH2017_108 and solution of the right–dihedra method. (c) Stereographic projection and rose diagram of faults and fractures of the sedimentary basement measured in outcrop LH2017_115. (d) Mafic dike cutting stratified limestone in outcrop LH2017_101 (photograph courtesy of K. Bär). (e) Vertical mafic sill intruded along the bedding planes of Mesozoic limestone in outcrop LH2018_02. (f) Rose diagram of mafic dikes and sills intruded in the sedimentary basement.

**Figure 4**: (a) Volcanotectonic map of the Los Humeros geothermal field and Los Potreros caldera area, on a shaded relief image obtained from the 1 m resolution DEM (illuminated from the E). Rose diagrams show the strike of faults and fractures measured in the field within the northern (light purple) and southern (light yellow) resurgent sectors. LH F.: Los Humeros Fault. Location of outcrops in Fig. 5, H43 geothermal well, 16 August 2015 and 08 February 2016 earthquake epicenters and C-C'-C'' topographic profile of Fig. 4b are shown. (b) E-W/NW-SE topographic profile along the C-C'-C'' trace shown in Fig. 4a.

**Figure 5**: Photographs and stereographic projections of fault and fracture data collected in the LHVC. Location in Fig. 4a. (a) Outcrop PDL54. NNW-SSE-striking Los Humeros fault with dip-slip normal slickenlines in hydrothermally altered pyroclastic fall deposit, stereographic projection of the fault data measured in the outcrop, and solution of the right–dihedra method. (b) Outcrop LH2017-36. NNE-SSW-striking normal fault in hydrothermally altered pyroclastic fall deposit showing the thermal signal of hot fluids circulating along the fault plane, stereographic projection of the fault data measured in the outcrop, and solution of the right–dihedra method (thermal image courtesy of C. Rochelle). (c) Outcrop PDL52. N-S-striking reverse fault in hydrothermally altered pyroclastic fall deposit along the curved trace of the Arroyo Grande fault, stereographic projection of the fault data measured in the outcrop, and solution of the right–dihedra method. (d) Outcrop LH2017-43. Rectilinear NNW-SSE scarp displacing a Holocene pahoehoe lava flow field.

**Figure 6**: (a) simplified volcanotectonic map with the location of geothermal wells shown in Fig. 6b and the traces of the geological-cross sections shown in Fig. 6c and 6d (full extension of the geological-cross sections traces is shown in Fig. 1). (b) Stratigraphic framework of LHVC and its basement (modified from Norini et al., 2015) and interpreted lithological well logs based on Unconformity Bounded Stratigraphic Units (UBSUs). (c) A-A' and (d) B-B' schematic geological cross-sections showing the subsurface geometry of the main structures and stratigraphic units, based on the geological map, structural field data, lithological well logs and geophysical and seismological data discussed in the text. Traces of the geological cross-sections are shown in Figs. 1 and 6a.

**Figure 7**: FMI log data recorded in H43 geothermal well (for location see Figs. 4a and 8a). (a) Rose diagram of fault planes imaged by FMI. (b) Rose diagram of IFs with measured depth (TD) < 1500 m and interpreted maximum horizontal stress in the upper part of the geothermal field. (c) Rose diagram of IFs with TD > 1500 m and interpreted maximum horizontal stress in the lower part of the geothermal field. (d) perspective view from SE of H43 well around 1588 m TD, with a fault plane and IF imaged by FMI.

**Figure 8**: (a) focal mechanism solutions of the 16 August 2015 and 08 February 2016 earthquakes plotted on the volcanotectonic map (see Fig. 4a). In the inset, local seismic network operating during the 2015 and 2016 earthquakes (modified from Lermo et al., 2016). (b) Perspective view from SE of the focal mechanism solutions of the 16 August 2015 and 08 February 2016 earthquakes and H43 well (location of Fig. 7d is shown).

**Figure 9**: (a) MT-MT' magnetotelluric profile along the SW-NE trace shown in Fig. 1. (b) Perspective view from south of the MT-MT' profile and A-A' geological cross-section. (c) Perspective view from south of the A-A' geological cross-section and focal mechanism solution of the 08 February 2016 earthquake. Legend of the geological cross-section is shown in Figs. 6b and 6d. Legend of the focal mechanism solution is shown in Fig. 8b.

**Figure 10**: (a) Schematic structural interpretation of the LHVC and its basement. (b) Simplified volcanotectonic map of the Los Humeros geothermal field, showing resurgence faults longer than 1.5 km and orientation of the regional inherited Mexican Fold and Thrust Belt (MFTB) and TMVB tectonic structures. LH F.: Los Humeros Fault. (c) Schematic not to scale structural interpretation along the C-C'-C'' trace shown in Fig. 4. P: inferred pressure source inducing caldera resurgence. Color of faults as in Fig. 10b.

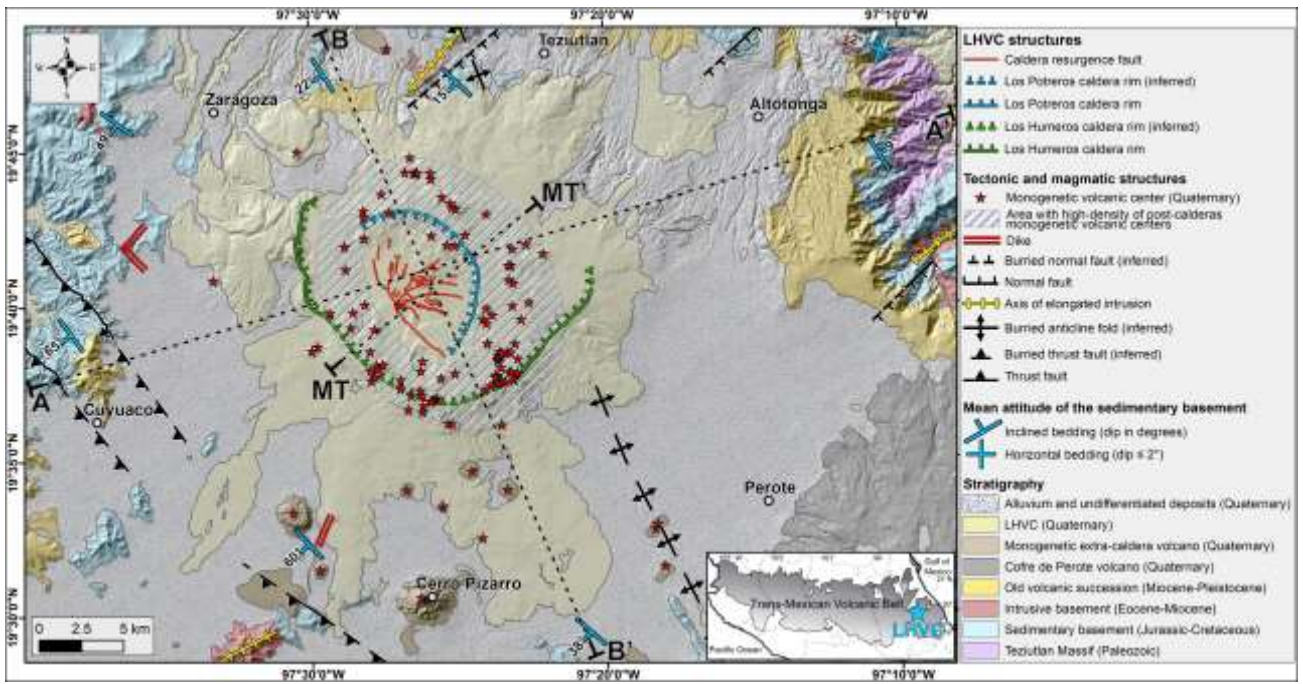

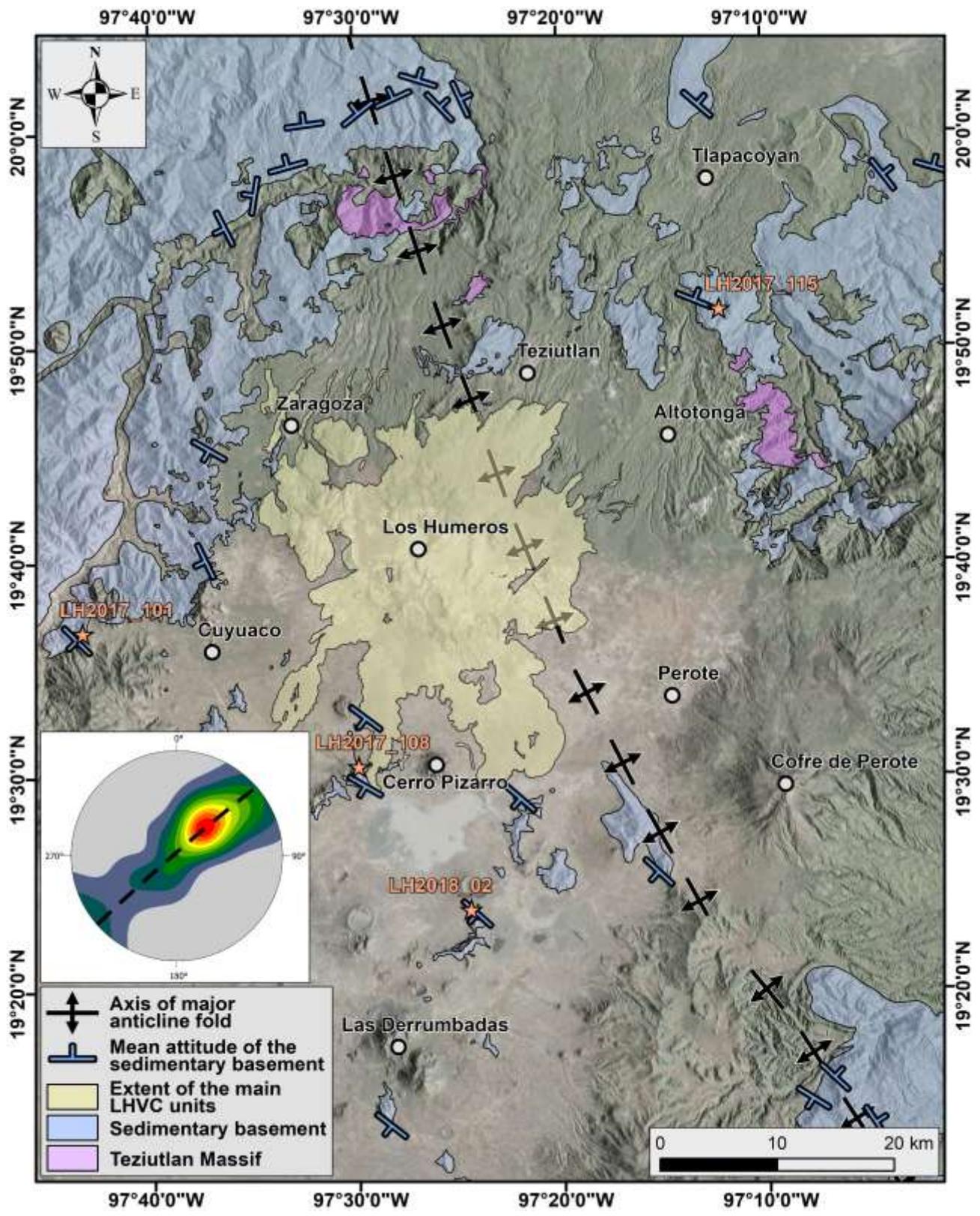

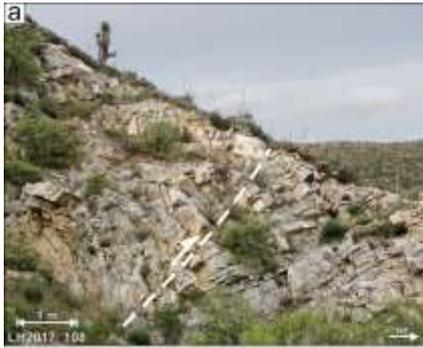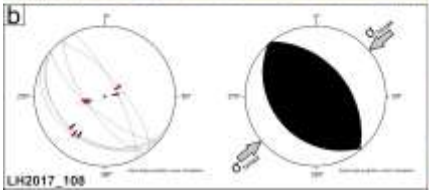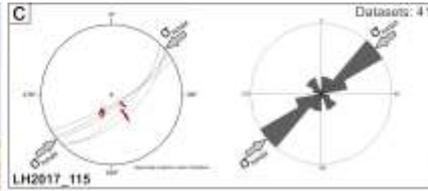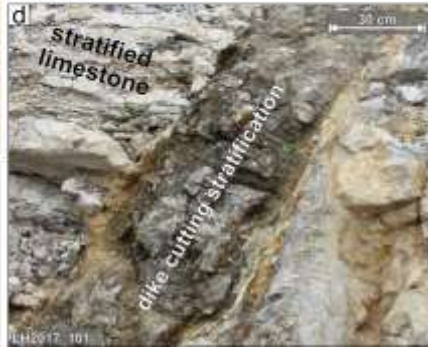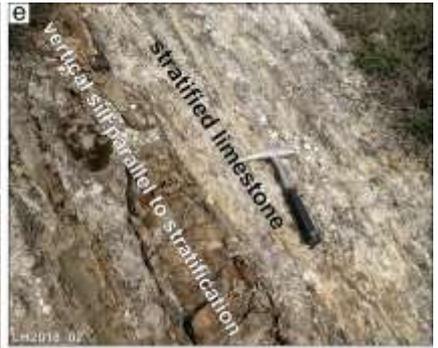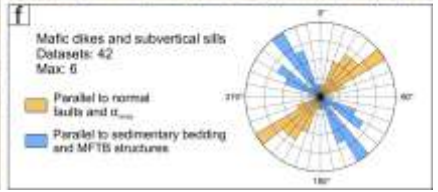

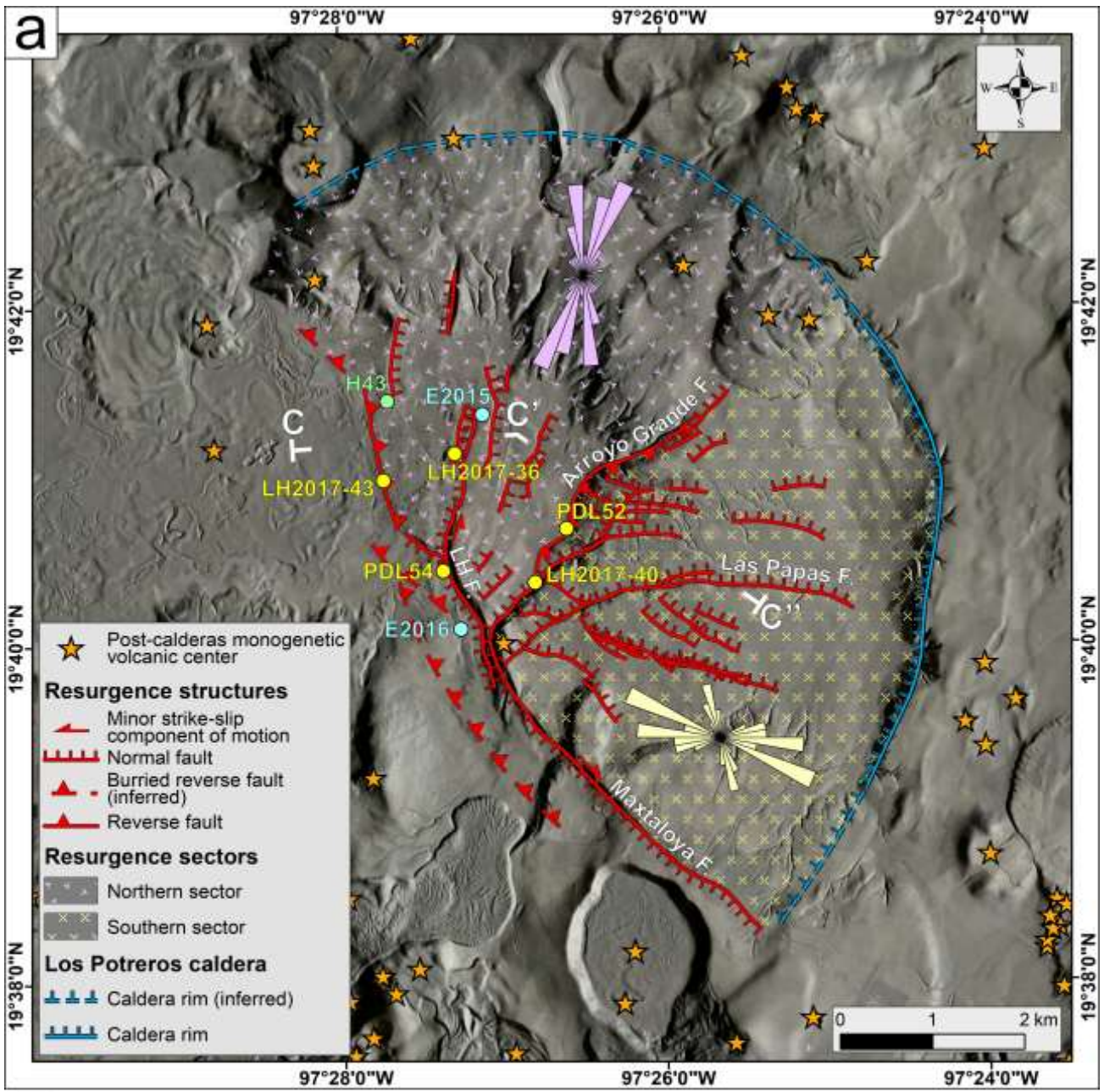
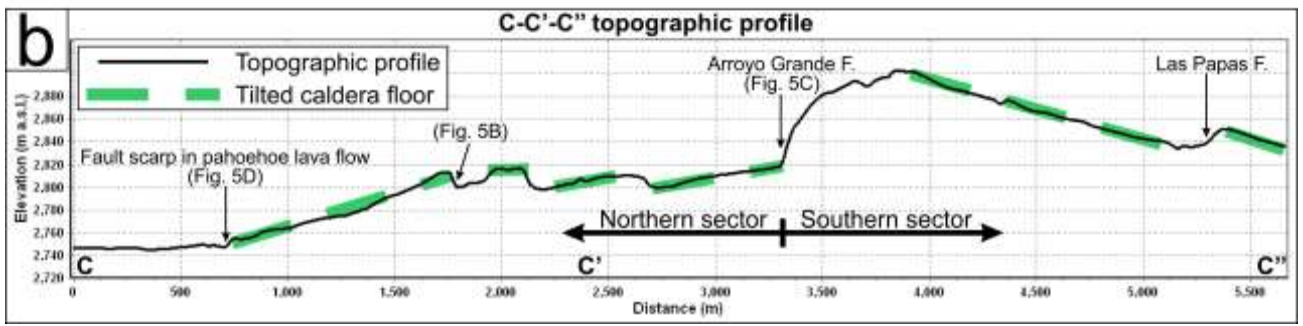

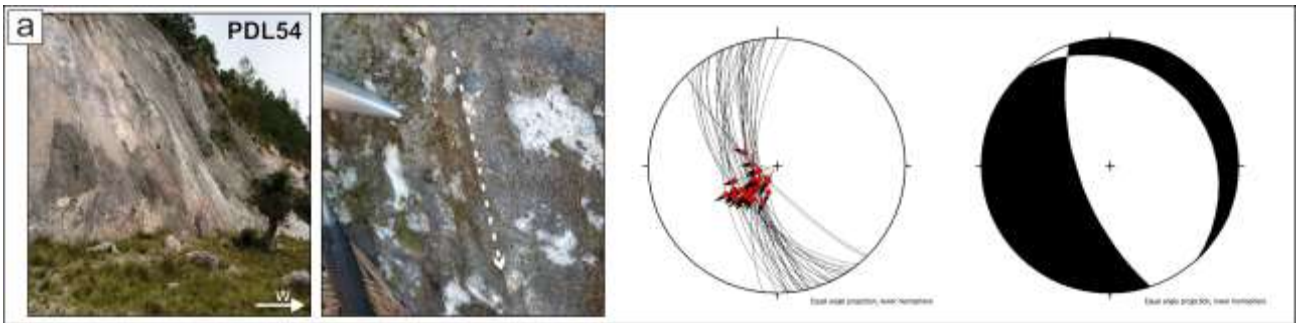
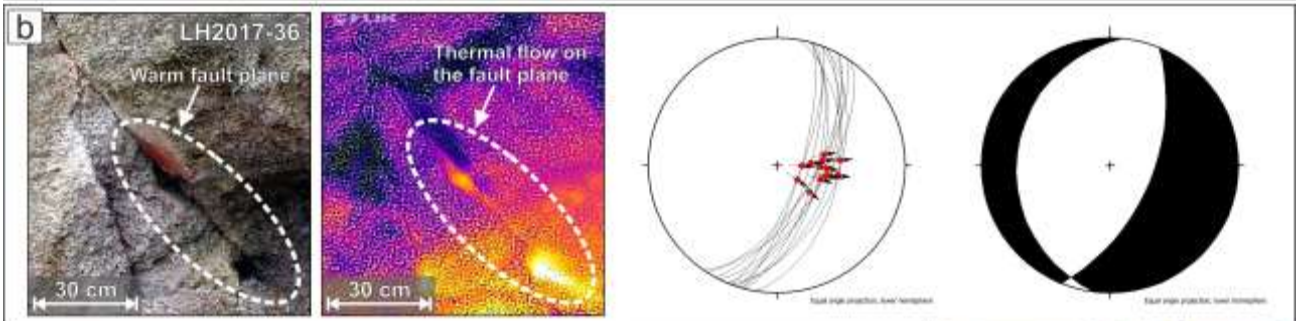
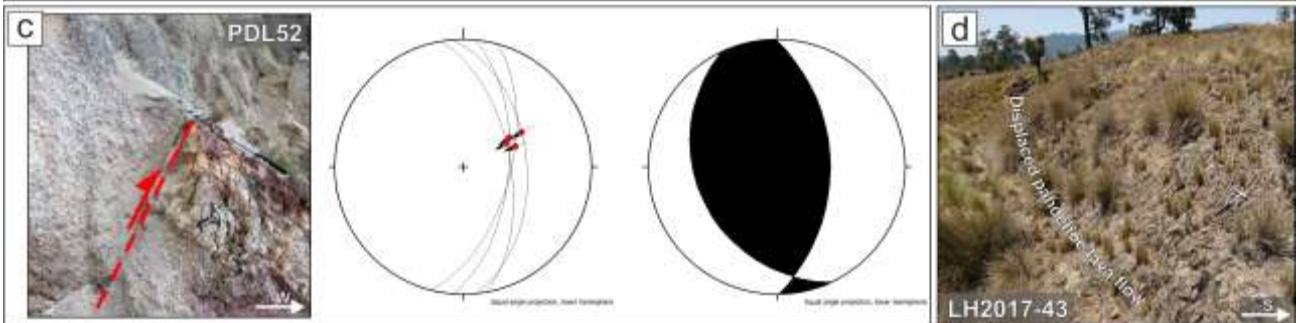
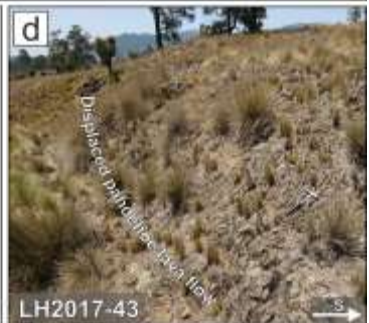
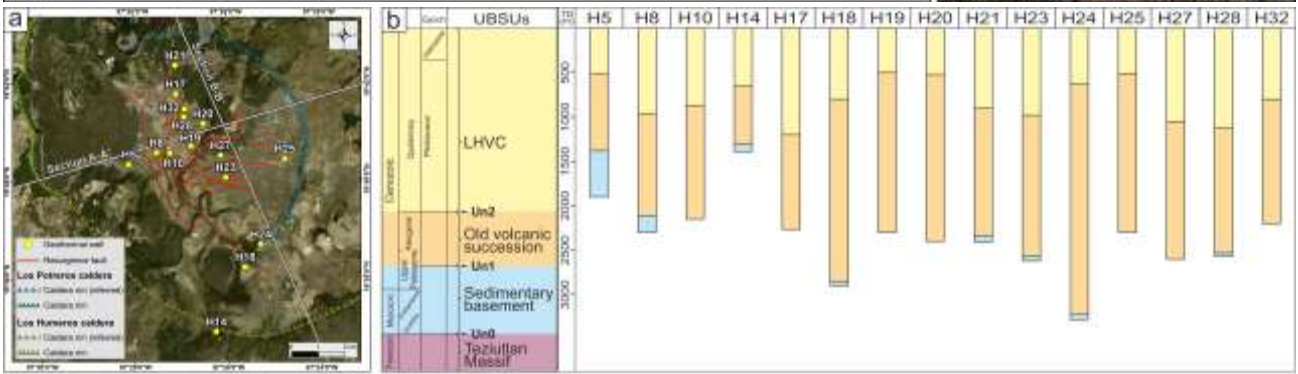
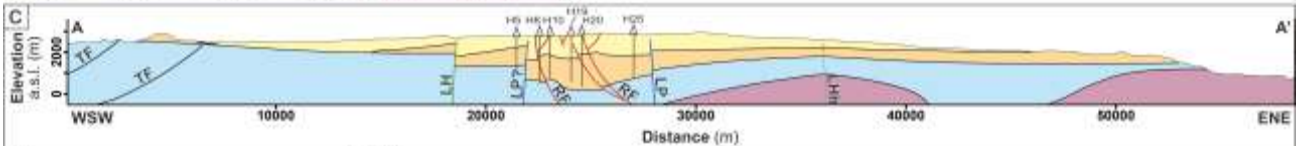
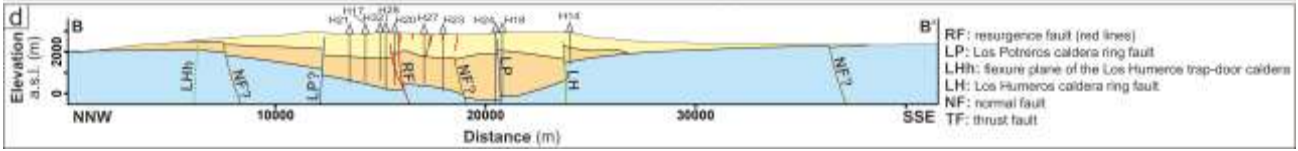

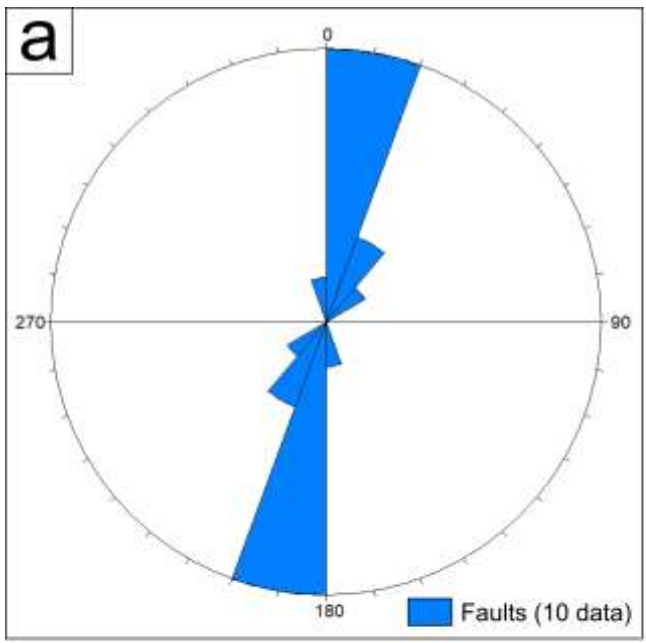
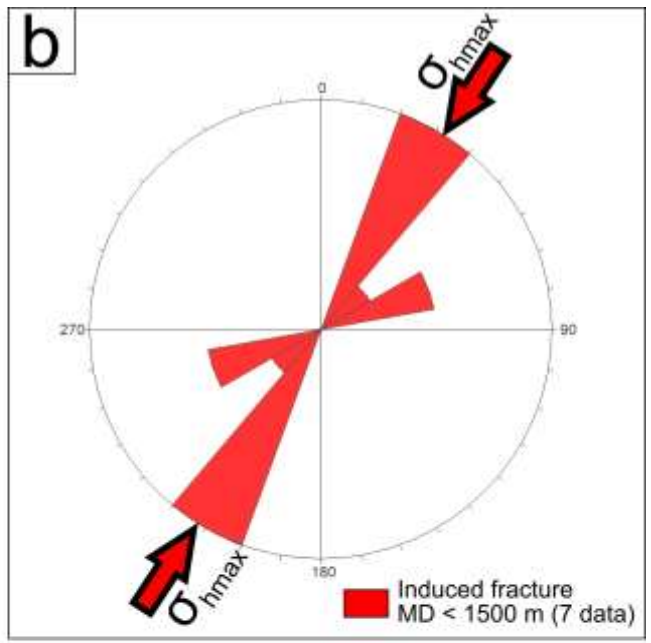
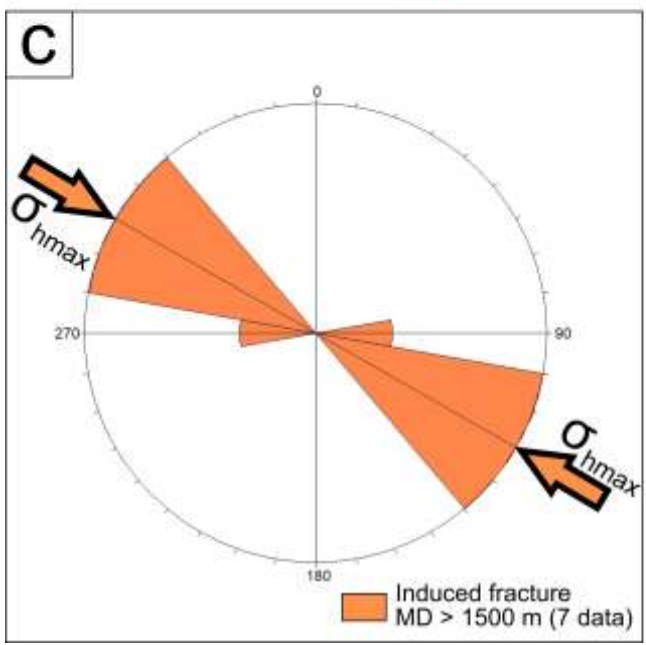
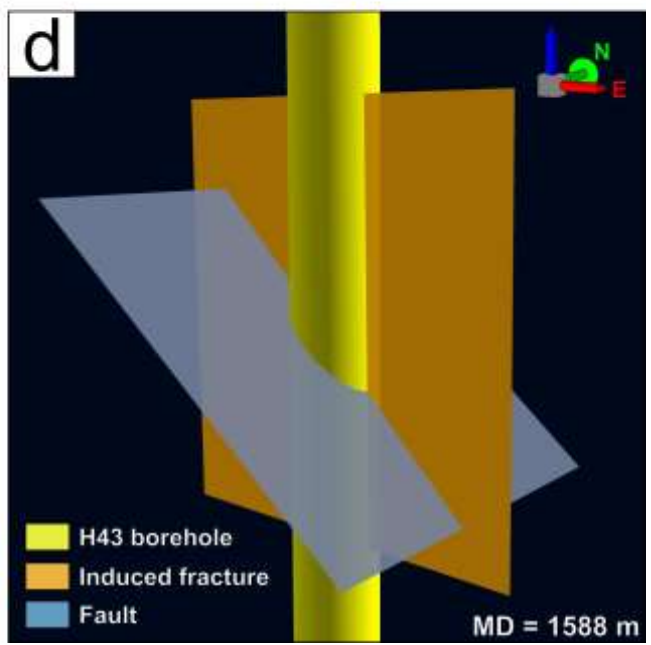

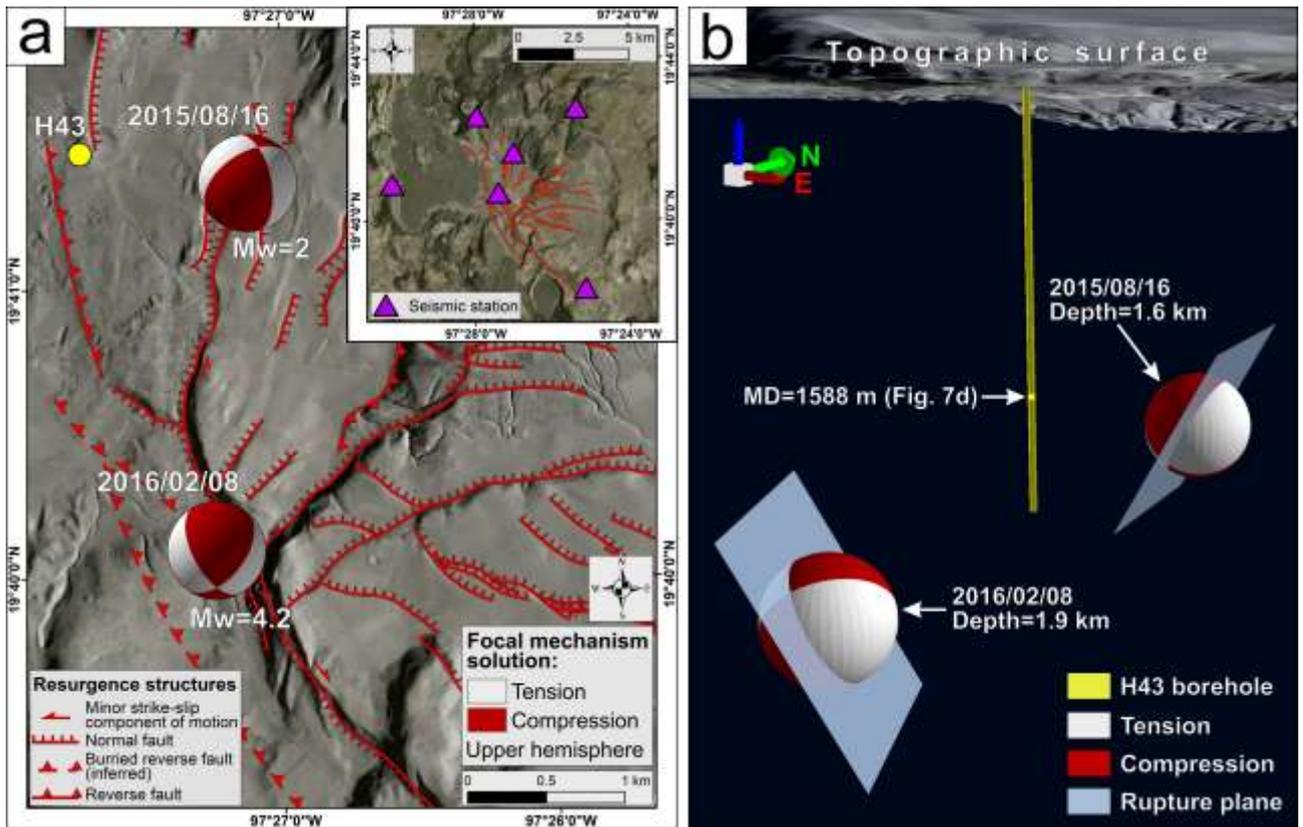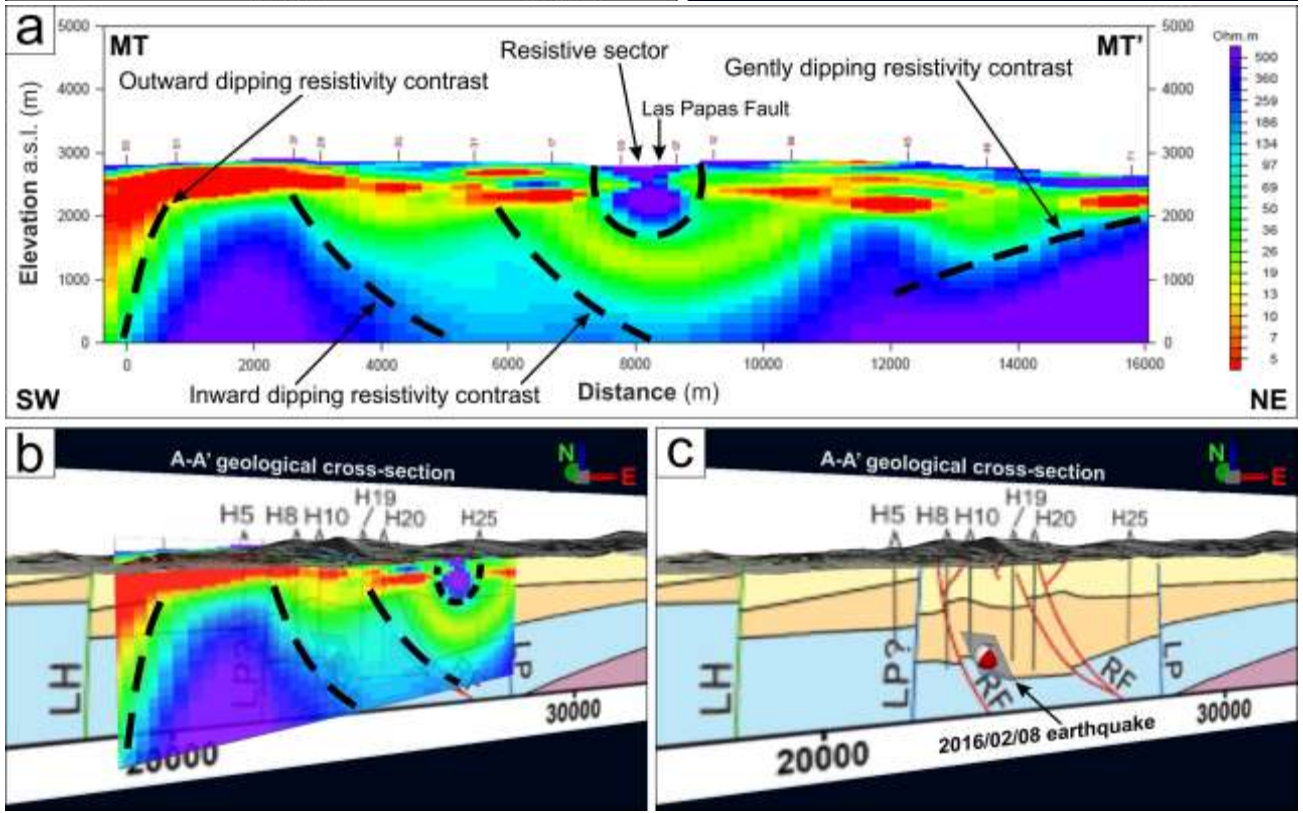

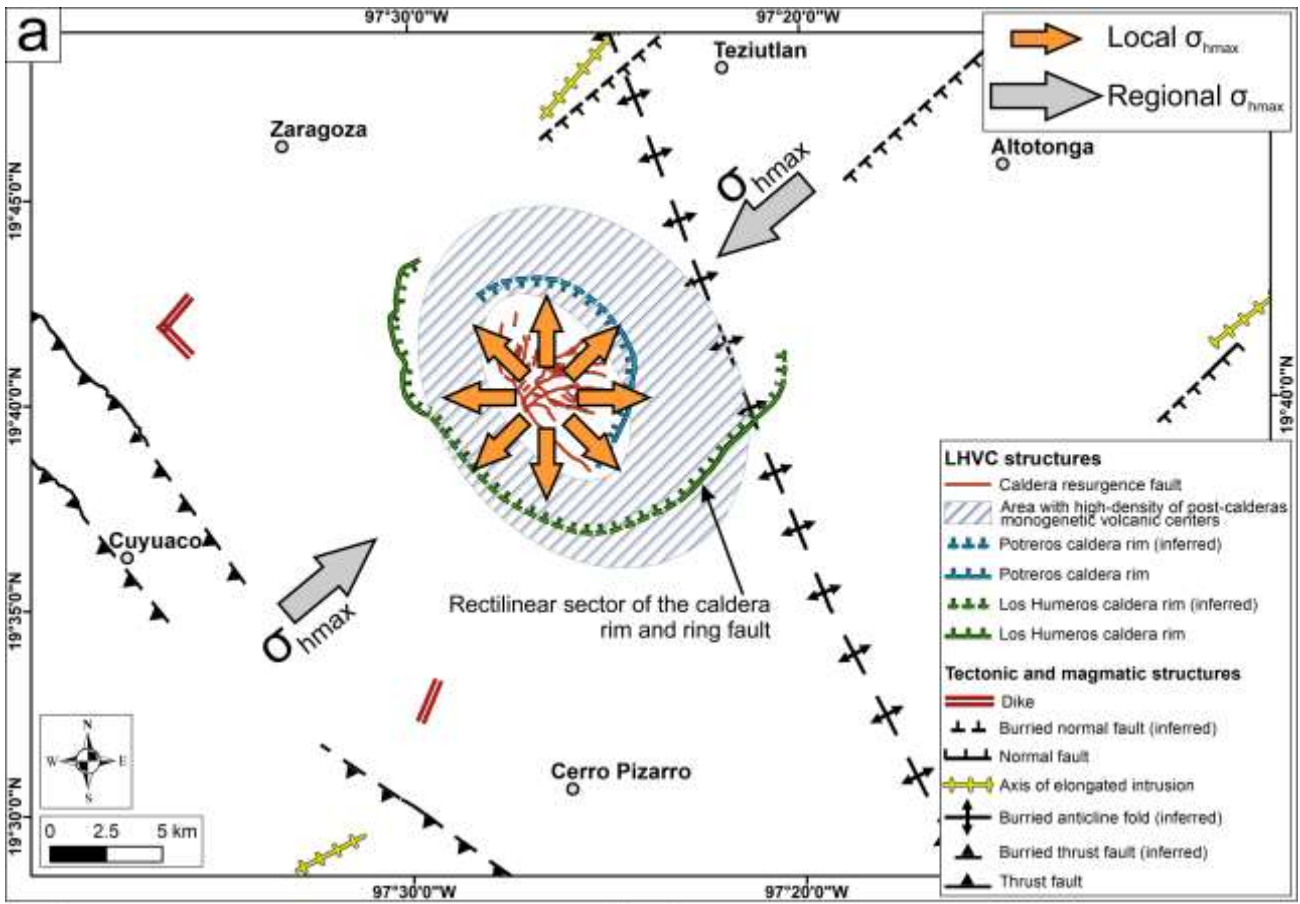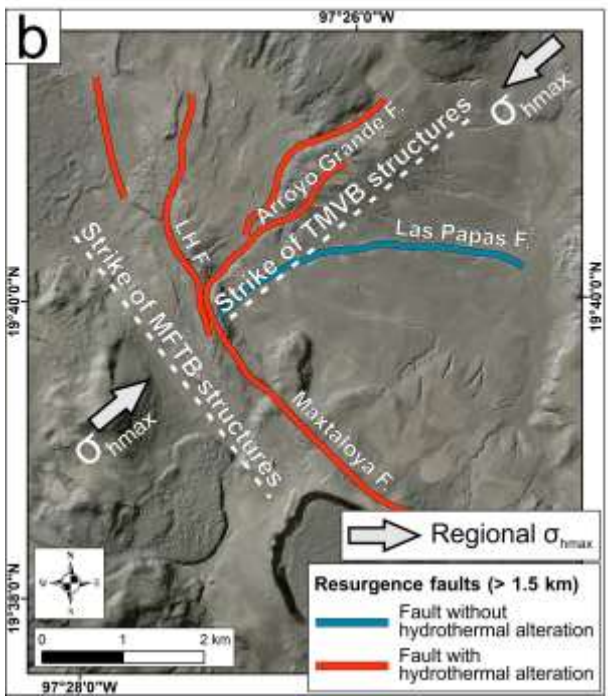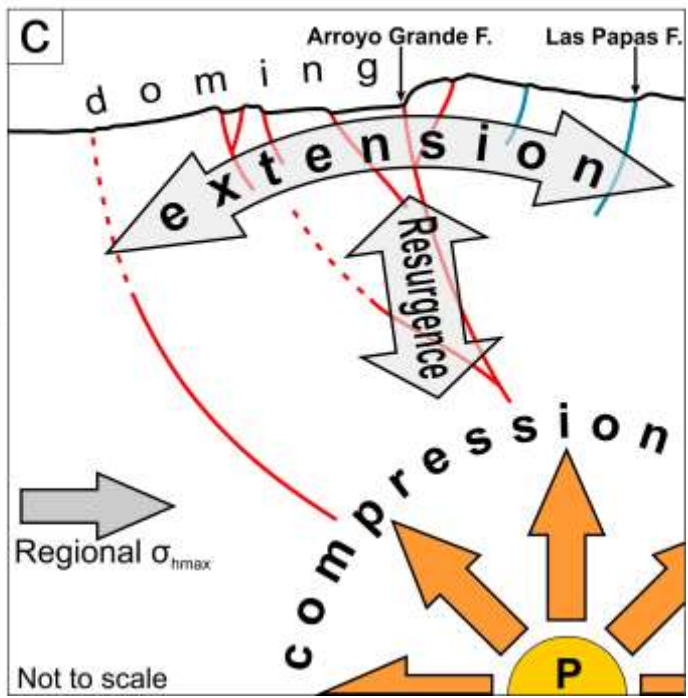